# The influence of various optimization algorithms on nuclear power plant steam turbine exergy efficiency and destruction

Vedran Mrzljak, Nikola Anđelić, Ivan Lorencin, Sandi Baressi Šegota

Faculty of Engineering, University of Rijeka, Vukovarska 58, 51000 Rijeka, Croatia
email: vedran.mrzljak@riteh.hr, nandelic@riteh.hr, ilorencin@riteh.hr, sbaressisegota@riteh.hr

**Abstract:** This paper presents an exergy analysis of the whole turbine, turbine cylinders and cylinder parts in four different operating regimes. Analyzed turbine operates in nuclear power plant while three of four operating regimes are obtained by using optimization algorithms - SA (Simplex Algorithm), GA (Genetic Algorithm) and IGSA (Improved Genetic-Simplex Algorithm). IGSA operating regime gives the highest developed mechanical power of the whole turbine equal to 1022.48 MW, followed by GA (1020.06 MW) and SA (1017.16 MW), while in Original operating regime whole turbine develop mechanical power equal to 996.29 MW. In addition, IGSA causes the highest increase in developed mechanical power of almost all cylinders and cylinder parts in comparison to the Original operating regime. All observed optimization algorithms increases the exergy destruction of the whole turbine in comparison to Original operating regime - the lowest increase causes IGSA, followed by GA and finally SA. The highest exergy efficiency of the whole turbine, equal to 85.92% is obtained by IGSA, followed by GA (85.89%) and SA (85.82%), while the lowest exergy efficiency is obtained in Original operating regime (85.70%). Analyzed turbine, which operates by using wet steam is low influenced by the ambient temperature change. IGSA, which shows dominant performance in exergy analysis parameters of the analyzed turbine, in certain situations is overpowered by GA. Therefore, in optimization of steam turbine performance, IGSA and GA can be recommended.

## 1 Introduction

The dominant usage of steam turbines nowadays is production of mechanical power and electric generators drive [1, 2]. Such steam turbines can be found in conventional [3, 4], nuclear [5, 6], combined or cogeneration [7-9], marine [10-12], or various other power plants [13-15]. Also, steam turbines can be used for various other purposes instead of electric generator drive. Steam turbines can be used in various power plants for the drive of pumps or ventilators [16-18], in marine steam power plants steam turbines are also used for main propulsion propeller drive [19, 20], in many complex power plants steam turbines can be used for the drive of various equipment [21, 22], etc.

Steam turbine operation in various power plants strongly differs. They can dominantly operate by using wet or superheated steam [23, 24], they can be composed of one or more turbine cylinders [25, 26], turbine cylinders can be single flow or dual flow (while dual flow cylinders can be symmetrical or asymmetrical) [27], turbine cylinders can be connected to the same or to the various shafts [28, 29], steam turbines are nowadays dominantly composed of many stages (however, still in exploitation can be found steam turbines with only one stage used for various auxiliary purposes) [30, 31], turbine stages can be of impulse, reaction or Curtis type [32], temperature and pressure of steam delivered to any steam turbine can significantly vary [33, 34]. Therefore, observations which pointed out that any steam turbine operates in a same manner is valid only in the most general and overall form.

In the scientific and professional literature can be found a variety of different techniques for steam turbines analysis and optimization. Although some researchers still perform various types of energy analyses related to steam turbines [35, 36], the dominance over energy analysis has exergy analysis [37-39] which takes into consideration the parameters of the ambient where the observed steam turbine operates, along with a fact that exergy analysis can be a proper baseline for the economic analysis, not only for steam turbines, but also for the entire power plants [40, 41]. Either energy or exergy analyses did not take into consideration steam turbine (or any other observed volume) inner structure [42, 43]. Along with mentioned, there also exist various other types of steam turbine numerical analyses which may (or may not) take into consideration turbine inner structure [44, 45].

Optimization of steam turbines (or the entire power plants) is nowadays often performed by using many optimization algorithms [46, 47]. Each optimization algorithm can be used not only for one specific problem, it is also applicable to a variety of engineering problems and challenges [48, 49]. Along with optimization algorithms, many researchers use various artificial intelligence methods and processes for improving the operation of steam turbines as an independent components [50, 51] or for improving the operation of entire power plants [52, 53]. Therefore, optimization and improvement possibilities for any observed component or a system are notable at the moment.

In this paper is investigated how the nuclear power plant operating parameters obtained by different optimization algorithms influenced exergy analysis of the whole steam turbine, turbine cylinders and each cylinder part and what are the differences in comparison to Original plant operating regime (standard plant operation). The main aim of all optimization algorithms was to increase overall nuclear power plant efficiency, but it was interesting to investigate how the same algorithms influenced exergy losses and exergy efficiencies of whole steam turbine, turbine cylinders and cylinder parts. Also, it is investigated did the ambient temperature change influenced steam turbine from nuclear power plant (which dominantly operate by using wet steam) more than turbine which dominantly operate by using superheated steam (turbine from conventional or marine steam power plant). Obtained results can be useful in the analysis and optimization of any complex steam turbine, because they show the influence of optimization algorithms not only on the whole turbine, but also on its cylinders and cylinder parts.

## 2 Description and operating characteristics of the analyzed steam turbine from nuclear power plant

Scheme of the analyzed steam turbine from nuclear power plant along with marked operating points required for the exergy analysis is presented in Fig. 1.

Steam turbine is a complex one and consists of four cylinders (one High Pressure Cylinder – HPC and three Low Pressure Cylinders – LPC). All turbine cylinders are of dual flow type - steam enters in the middle of the cylinder and simultaneously expands through left and right cylinder part. All low pressure cylinders are symmetrical ones – steam extractions at both cylinder parts has the same pressure, temperature and mass flow rate. LPC1 and LPC3 are actually identical cylinders, while LPC2 differs from the other low pressure cylinders due to different steam temperature, pressure and mass flow rate extracted in the first extraction.

HPC is asymmetrical cylinder – steam extraction in both of its parts occurs at different pressure and temperature, while the extracted steam mass flow rate is also different.

As can be seen from Fig. 1, all four steam turbine cylinders are connected to the same shaft which drives an electric generator. Such arrangement of turbine cylinders is standard for nuclear power plants – in comparison to conventional steam power plants, nuclear power plants did not posses intermediate pressure cylinder [54, 55].

The whole steam turbine process begins from steam delivery to HPC from steam generators (the dominant steam mass flow rate from steam generators is delivered to HPC, while a small part of the steam mass flow rate is delivered to Re-Heater – RH2), Fig. 1. The steam expands through left and right HPC parts (both HPC extractions deliver a certain steam mass flow rate to high pressure feed water heaters [56, 57]). After expansion in HPC, one small amount of steam mass flow rate is delivered to the deaerator, while the dominant steam mass flow rate is re-heated - in nuclear power plants, steam re-heating is performed firstly through Steam Separator (SS) which removes the dominant amount of water droplets from wet steam and two Re-Heaters (RH1 and RH2) which increases steam temperature (by using steam of higher temperature for that purpose). Therefore, steam re-heating process in nuclear power plant notable differs from steam re-heating process in conventional steam power plants (because in nuclear power plants fossil fuels are not used at all) [58, 59].

After steam re-heating process, one small amount of steam is delivered to the low power steam turbine which drives the main feed water pump [17], while other steam mass flow rate is equally distributed to all three low pressure cylinders. Steam extractions of all low pressure cylinders lead steam to low pressure condensate heaters [60, 61]. In Fig. 1 numeration of low pressure condensate heaters is performed in a manner that heater number increases from the main steam condenser to deaerator, while high pressure feed water heater number increases from the deaerator to steam generators. Remaining steam mass flow rate, after expansion in each part of each LPC is delivered to the main steam condenser for condensation [62, 63].

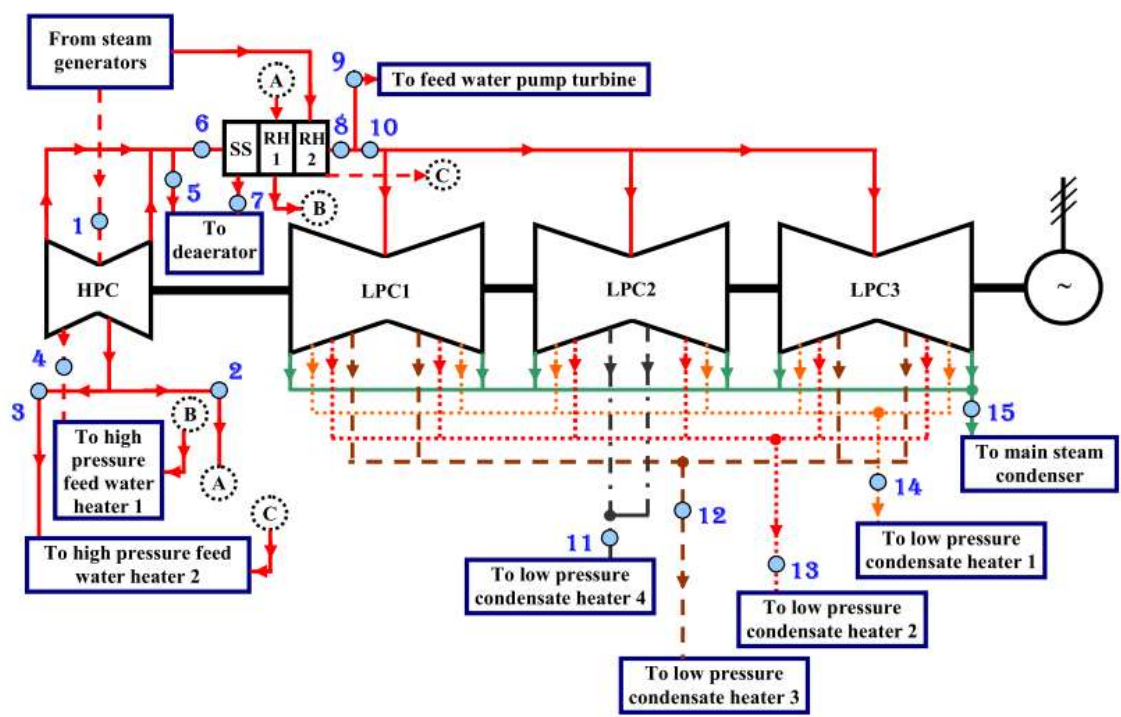

**Fig. 1** Scheme of the analyzed steam turbine from nuclear power plant along with marked operating points required for the exergy analysis
**Source:** Authors

## 3 Exergy analysis equations

Thermodynamic performance is best evaluated using exergy analysis because it provides more insights and is more useful in efficiency-improvement efforts than energy analysis alone [64, 65]. For any observed control volume, exergy analysis shows performance in relation to the environment [66, 67], while energy analysis did not have any relation with the environment in which control volume operates [68-70]. Therefore, in the scientific and professional literature can be found that many researchers selected exergy analysis instead of energy analysis for researching the operation of various control volumes or systems [71-73]. Also, in this paper was most proper and adequate to analyze exergy efficiencies and losses of the observed steam turbine, its cylinders and cylinder parts and investigate the influences of all observed optimization algorithms.

## 3.1 General exergy equations and balances

In the exergy analysis exists a few general equations and balances which must be always satisfied, regardless of the observed control volume or its operating regime [74-76].

First of all general exergy analysis equations is the equation of exergy balance, which can be often found in the literature [77-79]. So, for any observed control volume, the equation of exergy balance is:

$$\dot{X}_{ex} + P_{in} + \sum \dot{E}x_{in} = P_{out} + \sum \dot{E}x_{out} + \dot{E}x_d. \qquad (1)$$

In the above equation, $P$ is mechanical power, $\dot{E}x_d$ is exergy destruction (exergy loss), index in is related to the inlet (input), while index out is related to the outlet (output) of the observed control volume. A total exergy flow of operating fluid ($\dot{E}x$) can be defined by using equation [80, 81]:

$$\dot{E}x = \dot{m} \cdot \varepsilon, \qquad (2)$$

where $\dot{m}$ is operating fluid mass flow rate, while $\varepsilon$ is specific exergy. The specific exergy calculation of any operating fluid stream can be performed by using equation [82]:

$$\varepsilon = (h - h_0) - T_0 \cdot (s - s_0), \qquad (3)$$

where $T$ is temperature, $h$ is specific enthalpy, $s$ is specific entropy and index 0 represents operating parameters of any fluid stream at the ambient state (at the ambient temperature and pressure). Therefore, Eq. 3 represents the direct involvement of the ambient state parameters into the exergy analysis of any observed control volume [83]. Also, Eq. 3 allows the change of the ambient temperature and pressure, so any observed control volume can be investigated in the whole range of ambient states (inside which the control volume operation can be expected) [84].

The last undefined parameter from Eq. 1 is the exergy transfer by heat ($\dot{X}_{ex}$) at the temperature $T$, which can be calculated by using an equation [85]:

$$\dot{X}_{ex} = \sum (1 - \frac{T_0}{T}) \cdot \dot{Q}, \qquad (4)$$

where $\dot{Q}$ is the energy transfer by heat. Also, for any control volume should be valid mass flow rate balance [86]:

$$\sum \dot{m}_{IN} = \sum \dot{m}_{OUT}. \qquad (5)$$

In the analyses of some control volumes, as for example in [25, 36, 87], mass flow rate leakages can be taken into account, but mass flow rate balance must always be satisfied because any mass flow rate cannot rise or disappear by itself.

Final general equation related to exergy analysis is definition of any control volume exergy efficiency. The exergy efficiency definition depends on the characteristics and operating principles of a control volume, so the exact equation may significantly vary. General representation of control volume exergy efficiency can be shown, according to [88, 89], as:

$$\eta_{ex} = \frac{\text{exergy output}}{\text{exergy input}}. \qquad (6)$$

## 3.2 Equations for the exergy analysis of the observed steam turbine, turbine cylinders and cylinder parts

All the equations related to exergy analysis of whole observed steam turbine, turbine cylinders and cylinder parts are developed in accordance to available literature [90-94], by using operating points presented in Fig. 1. The equations presented in this subsection are identical in all turbine observed operating regimes (Original, IGSA, GA and SA) at each observed ambient temperature – the differences occur in steam operating parameters for each operating regime.

The essential parameter required in the exergy analysis of any steam turbine is developed mechanical power ($P$). Developed mechanical power of the whole turbine, turbine cylinders and cylinder parts is calculated by using steam operating parameters and corresponding equations are presented in Table 1.

**Table 1** Equations for mechanical power calculation – whole turbine, turbine cylinders and cylinder parts

| Component* | Mechanical power | Eq. |
|---|---|---|
| **HPC-L** | $P_{HPC-L} = \frac{\dot{m}_1}{2} \cdot (h_1 - h_4) + \left(\frac{\dot{m}_1}{2} - \dot{m}_4\right) \cdot (h_4 - h_5)$ | (7) |
| **HPC-R** | $P_{HPC-R} = \frac{\dot{m}_1}{2} \cdot (h_1 - h_2) + \left(\frac{\dot{m}_1}{2} - \dot{m}_2 - \dot{m}_3\right) \cdot (h_2 - h_5)$ | (8) |
| **HPC-EC** | $P_{HPC-EC} = P_{HPC-L} + P_{HPC-R}$ | (9) |
| **LPC1-L** | $P_{LPC1-L} = \frac{\dot{m}_{10}}{6} \cdot (h_{10} - h_{12}) + \left(\frac{\dot{m}_{10}}{6} - \frac{\dot{m}_{12}}{4}\right) \cdot (h_{12} - h_{13})$ $+ \left(\frac{\dot{m}_{10}}{6} - \frac{\dot{m}_{12}}{4} - \frac{\dot{m}_{13}}{6}\right) \cdot (h_{13} - h_{14})$ $+ \left(\frac{\dot{m}_{10}}{6} - \frac{\dot{m}_{12}}{4} - \frac{\dot{m}_{13}}{6} - \frac{\dot{m}_{14}}{6}\right) \cdot (h_{14} - h_{15})$ | (10) |
| **LPC1-R** | $P_{LPC1-R} = P_{LPC1-L}$** | (11) |
| **LPC1-EC** | $P_{LPC1-EC} = P_{LPC1-L} + P_{LPC1-R}$ | (12) |
| **LPC2-L** | $P_{LPC2-L} = \frac{\dot{m}_{10}}{6} \cdot (h_{10} - h_{11}) + \left(\frac{\dot{m}_{10}}{6} - \frac{\dot{m}_{11}}{2}\right) \cdot (h_{11} - h_{13})$ $+ \left(\frac{\dot{m}_{10}}{6} - \frac{\dot{m}_{11}}{2} - \frac{\dot{m}_{13}}{6}\right) \cdot (h_{13} - h_{14})$ $+ \left(\frac{\dot{m}_{10}}{6} - \frac{\dot{m}_{11}}{2} - \frac{\dot{m}_{13}}{6} - \frac{\dot{m}_{14}}{6}\right) \cdot (h_{14} - h_{15})$ | (13) |
| **LPC2-R** | $P_{LPC2-R} = P_{LPC2-L}$** | (14) |
| **LPC2-EC** | $P_{LPC2-EC} = P_{LPC2-L} + P_{LPC2-R}$ | (15) |
| **LPC3-L** | $P_{LPC3-L} = P_{LPC1-L}$*** | (16) |
| **LPC3-R** | $P_{LPC3-R} = P_{LPC1-R}$*** | (17) |
| **LPC3-EC** | $P_{LPC3-EC} = P_{LPC1-EC}$*** | (18) |
| **WT** | $P_{WT} = P_{HPC-EC} + P_{LPC1-EC} + P_{LPC2-EC} + P_{LPC3-EC}$ | (19) |

\* L = Left part; R = Right part; EC = Entire Cylinder; WT = Whole Turbine
\*\* Symmetry of both cylinder parts
\*\*\* LPC1 and LPC3 are identical cylinders (the same steam mass flow rates, pressures and temperatures)
**Source:** Authors

Equations for exergy destruction calculation of the whole turbine, turbine cylinders and cylinder parts are presented in Table 2.

In the calculation of exergy destructions, special attention should be pointed to the steam mass flow rates at the HPC outlet (HPC is asymmetrical cylinder, so steam mass flow rates at the outlets of HPC-L and HPC-R are not equal). As mentioned before, LPC1 and LPC3 are identical cylinders, therefore steam mass flow rates at the outlet of each cylinder parts are identical. However, steam mass flow rates extracted from LPC2 differ (in comparison to LPC1 and LPC3), so steam mass flow rates at the outlet of both LPC2 parts are identical, but they differ in comparison to LPC1 and LPC3 outlet streams. If the explained facts related to outlet steam mass flow rates from the cylinders were not taken into consideration, it can easily lead to obvious mistakes which significantly influenced the results of the entire exergy analysis. Mentioned outlet steam mass flow rates are also important in the exergy destruction calculation, so the special attention is pointed out on its correct definition, as can be seen in Table 2.

**Table 2** Equations for the exergy destruction calculation – whole turbine, turbine cylinders and cylinder parts

| Component* | Exergy destruction | Eq. |
|---|---|---|
| **HPC-L** | $\dot{E}x_{d,HPC-L} = \frac{\dot{m}_1}{2} \cdot \varepsilon_1 - \dot{m}_4 \cdot \varepsilon_4 - \left(\frac{\dot{m}_1}{2} - \dot{m}_4\right) \cdot \varepsilon_5 - P_{HPC-L}$ | (20) |
| **HPC-R** | $\dot{E}x_{d,HPC-R} = \frac{\dot{m}_1}{2} \cdot \varepsilon_1 - \dot{m}_2 \cdot \varepsilon_2 - \dot{m}_3 \cdot \varepsilon_3 - \left(\frac{\dot{m}_1}{2} - \dot{m}_2 - \dot{m}_3\right) \cdot \varepsilon_5 - P_{HPC-R}$ | (21) |
| **HPC-EC** | $\dot{E}x_{d,HPC-EC} = \dot{E}x_{d,HPC-L} + \dot{E}x_{d,HPC-R}$ | (22) |
| **LPC1-L** | $\dot{E}x_{d,LPC1-L} = \frac{\dot{m}_{10}}{6} \cdot \varepsilon_{10} - \frac{\dot{m}_{12}}{4} \cdot \varepsilon_{12} - \frac{\dot{m}_{13}}{6} \cdot \varepsilon_{13} - \frac{\dot{m}_{14}}{6} \cdot \varepsilon_{14} - \left(\frac{\dot{m}_{10}}{6} - \frac{\dot{m}_{12}}{4} - \frac{\dot{m}_{13}}{6} - \frac{\dot{m}_{14}}{6}\right) \cdot \varepsilon_{15} - P_{LPC1-L}$ | (23) |
| **LPC1-R** | $\dot{E}x_{d,LPC1-R} = \dot{E}x_{d,LPC1-L}$** | (24) |
| **LPC1-EC** | $\dot{E}x_{d,LPC1-EC} = \dot{E}x_{d,LPC1-L} + \dot{E}x_{d,LPC1-R}$ | (25) |
| **LPC2-L** | $\dot{E}x_{d,LPC2-L} = \frac{\dot{m}_{10}}{6} \cdot \varepsilon_{10} - \frac{\dot{m}_{11}}{2} \cdot \varepsilon_{11} - \frac{\dot{m}_{13}}{6} \cdot \varepsilon_{13} - \frac{\dot{m}_{14}}{6} \cdot \varepsilon_{14} - \left(\frac{\dot{m}_{10}}{6} - \frac{\dot{m}_{11}}{2} - \frac{\dot{m}_{13}}{6} - \frac{\dot{m}_{14}}{6}\right) \cdot \varepsilon_{15} - P_{LPC2-L}$ | (26) |
| **LPC2-R** | $\dot{E}x_{d,LPC2-R} = \dot{E}x_{d,LPC2-L}$** | (27) |
| **LPC2-EC** | $\dot{E}x_{d,LPC2-EC} = \dot{E}x_{d,LPC2-L} + \dot{E}x_{d,LPC2-R}$ | (28) |
| **LPC3-L** | $\dot{E}x_{d,LPC3-L} = \dot{E}x_{d,LPC1-L}$*** | (29) |
| **LPC3-R** | $\dot{E}x_{d,LPC3-R} = \dot{E}x_{d,LPC1-R}$*** | (30) |
| **LPC3-EC** | $\dot{E}x_{d,LPC3-EC} = \dot{E}x_{d,LPC1-EC}$*** | (31) |
| **WT** | $\dot{E}x_{d,WT} = \dot{E}x_{d,HPC-EC} + \dot{E}x_{d,LPC1-EC} + \dot{E}x_{d,LPC2-EC} + \dot{E}x_{d,LPC3-EC}$ | (32) |

* L = Left part; R = Right part; EC = Entire Cylinder; WT = Whole Turbine
** Symmetry of both cylinder parts
*** LPC1 and LPC3 are identical cylinders (the same steam mass flow rates, pressures and temperatures)
**Source:** Authors

The exergy efficiency of the whole turbine, turbine cylinders and cylinder parts are calculated according to [95, 96] by using always the same equation:

$$\eta_{ex,(i)} = \frac{P_{(i)}}{\dot{E}x_{d,(i)} + P_{(i)}},\qquad(33)$$

where *i* denotes each component from Table 1 and Table 2 (each cylinder, cylinder part and the whole turbine).

In this analysis is also investigated the influence of the ambient temperature change on each observed turbine cylinder, cylinder part and the whole turbine in all four operating regimes. The analysis equations presented in this section remains the same, regardless of the observed ambient temperature, the changeable variable in all the equations is specific exergy of fluid streams.

## 4 Operating parameters required for the observed steam turbine exergy analysis

Operating parameters required for the exergy analysis of the whole observed steam turbine, its cylinders and cylinder parts are found in [97]. In mentioned literature the authors have made optimization of the whole nuclear power plant with an aim to increase the entire plant overall efficiency. Based on the measured results from nuclear power plant (Original operating regime), the authors have used three optimization algorithms: SA (Simplex Algorithm), GA (Genetic Algorithm) and IGSA (Improved Genetic-Simplex Algorithm). Each optimization algorithm provided a new operating parameters in each nuclear power plant operating point (mass flow rates, specific enthalpies, temperatures and pressures). During the optimization by using each of three optimization algorithms are respected operating limits of each component from the observed nuclear power plant (provided by manufacturers). The final conclusion in the described research was that all optimization algorithms increases overall power plant efficiency (in comparison to Original operating regime), while the highest overall efficiency is obtained by IGSA, followed by GA and finally SA.

As the mechanical power developed by the whole turbine is one of two essential elements which define overall plant efficiency, the authors of this paper was interested how the newly obtained operating parameters (by each optimization algorithm) in each power plant operating point influenced exergy efficiencies and destructions of each turbine cylinder, cylinder part and the whole turbine. Also, it was interesting to analyze did a new operating parameters obtained by each algorithm notably influenced steam turbine sensitivity on the ambient temperature change, especially by taking into consideration fact that the dominant operation of any steam turbine from any nuclear power plant is performed with wet (not superheated) steam. As presented in [98], components which operates by using wet steam (in this example, heat exchangers) can be notably influenced by the ambient temperature change. So, the analysis presented in this paper represents a continuation of a previous research with an aim to get a deeper insight into the steam turbine operation after optimization of the whole power plant.

Steam operating parameters of each stream related to the analyzed steam turbine (according to operating points presented in Fig. 1) in Original, IGSA, GA and SA operating regimes are presented in Table 3 and Table 4. In Table 3 are presented steam mass flow rates and specific enthalpies, while in Table 4 are presented steam pressures and temperatures. All the other operating parameters required for the exergy analysis are calculated from the presented ones by using NIST-REFPROP 9.0 software [99].

In the exergy analysis of any control volume the base ambient state can be set arbitrarily, so in the literature can be found various selections [100, 101]. In this analysis, the base ambient state is defined by selecting the ambient pressure of 1 bar and the ambient temperature of 25 °C.

**Table 3.** Steam mass flow rate and specific enthalpy in each operating point of the observed turbine [97]

| Operating point* | Mass flow rate (kg/s) | | | | Specific enthalpy (kJ/kg) | | | |
|---|---|---|---|---|---|---|---|---|
| | Original | IGSA | GA | SA | Original | IGSA | GA | SA |
| 1 | 1532.70 | 1587.40 | 1579.70 | 1571.80 | 2768.9 | 2757.5 | 2758.1 | 2759.4 |
| 2 | 72.01 | 86.41 | 83.07 | 77.51 | 2652.4 | 2632.9 | 2633.8 | 2635.8 |
| 3 | 76.48 | 77.02 | 72.12 | 78.77 | 2652.4 | 2632.9 | 2633.8 | 2635.8 |
| 4 | 111.02 | 133.22 | 133.17 | 124.79 | 2597.9 | 2579.5 | 2580.2 | 2582.1 |
| 5 | 64.70 | 51.99 | 59.54 | 56.76 | 2476.3 | 2467.0 | 2466.6 | 2467.4 |
| 6 | 1208.50 | 1239.00 | 1231.80 | 1234.00 | 2476.3 | 2467.0 | 2466.6 | 2467.4 |
| 7 | 165.80 | 181.50 | 180.30 | 179.70 | 714.2 | 735.4 | 733.3 | 730.8 |
| 8 | 1042.70 | 1057.40 | 1051.50 | 1054.30 | 2984.5 | 2979.5 | 2978.8 | 2976.5 |
| 9 | 31.10 | 31.10 | 31.10 | 31.10 | 2984.5 | 2979.5 | 2978.8 | 2976.5 |
| 10 | 1011.60 | 1026.30 | 1020.40 | 1023.20 | 2984.5 | 2979.5 | 2978.8 | 2976.5 |
| 11 | 41.42 | 33.14 | 34.84 | 35.95 | 2823.6 | 2814.0 | 2813.9 | 2813.6 |
| 12 | 40.24 | 48.25 | 46.50 | 46.18 | 2712.1 | 2703.1 | 2703.1 | 2702.0 |
| 13 | 48.36 | 58.03 | 53.87 | 55.68 | 2617.0 | 2609.1 | 2608.9 | 2607.9 |
| 14 | 52.23 | 62.67 | 59.27 | 58.48 | 2483.2 | 2476.7 | 2476.4 | 2475.4 |
| 15 | 829.35 | 824.20 | 825.90 | 826.90 | 2350.4 | 2334.5 | 2334.2 | 2335.0 |

\* Numeration of operating points is performed according to markings from Fig. 1.

**Table 4.** Steam pressure and temperature in each operating point of the observed steam turbine [97]

| Operating point* | Pressure (MPa) | | | | Temperature (°C) | | | |
|---|---|---|---|---|---|---|---|---|
| | Original | IGSA | GA | SA | Original | IGSA | GA | SA |
| 1 | 6.110 | 6.910 | 6.870 | 6.790 | 276.7 | 285.0 | 284.6 | 283.7 |
| 2 | 2.760 | 2.960 | 2.940 | 2.920 | 229.2 | 233.0 | 232.8 | 232.4 |
| 3 | 2.760 | 2.960 | 2.940 | 2.920 | 229.2 | 233.0 | 232.8 | 232.4 |
| 4 | 1.910 | 2.050 | 2.040 | 2.020 | 210.0 | 213.5 | 213.3 | 212.9 |
| 5 | 0.783 | 0.879 | 0.869 | 0.857 | 169.5 | 174.3 | 173.9 | 173.3 |
| 6 | 0.783 | 0.879 | 0.869 | 0.857 | 169.5 | 174.3 | 173.9 | 173.3 |
| 7 | 0.771 | 0.865 | 0.855 | 0.844 | 168.9 | 173.7 | 173.2 | 172.6 |
| 8 | 0.747 | 0.838 | 0.829 | 0.818 | 265.1 | 264.1 | 263.6 | 262.4 |
| 9 | 0.740 | 0.830 | 0.821 | 0.810 | 265.0 | 264.0 | 263.5 | 262.3 |
| 10 | 0.740 | 0.830 | 0.821 | 0.810 | 265.0 | 264.0 | 263.5 | 262.3 |
| 11 | 0.386 | 0.423 | 0.419 | 0.414 | 181.9 | 178.5 | 178.3 | 177.6 |
| 12 | 0.202 | 0.220 | 0.218 | 0.216 | 123.0 | 123.3 | 123.0 | 122.7 |
| 13 | 0.096 | 0.105 | 0.104 | 0.103 | 98.5 | 101.1 | 100.8 | 100.5 |
| 14 | 0.036 | 0.040 | 0.039 | 0.039 | 73.3 | 75.7 | 75.4 | 75.1 |
| 15 | 0.008 | 0.008 | 0.008 | 0.008 | 40.3 | 40.8 | 40.6 | 40.6 |

\* Numeration of operating points is performed according to markings from Fig. 1.

Steam turbines which operate in nuclear power plants strongly differs in relation to operating medium characteristics in comparison to steam turbines from conventional power plants. While in conventional steam power plants turbine operate dominantly by using superheated steam [102], steam turbines from nuclear power plants dominantly operate by using wet steam [103].

The same conclusion can be derived for the analyzed steam turbine, what is presented in Fig. 2. For the turbine analyzed in this paper the entire HPC (both of its parts) operates by using wet steam. All LPC's of the analyzed turbine has five characteristic operating points and in three of them steam is under the saturation line (wet steam), therefore it can be concluded that the dominant part of each LPC also operates with wet steam, Fig. 2. Wet steam consist of steam and a certain amount of water droplets which causes significant erosion on the turbine blades (erosion is much higher than in the turbines which dominantly operate by using superheated steam) [1, 32]. This is a reason why the maintenance must be much more rigid and why the turbine blades are changed much more often in nuclear power plants (in comparison to conventional steam power plants).

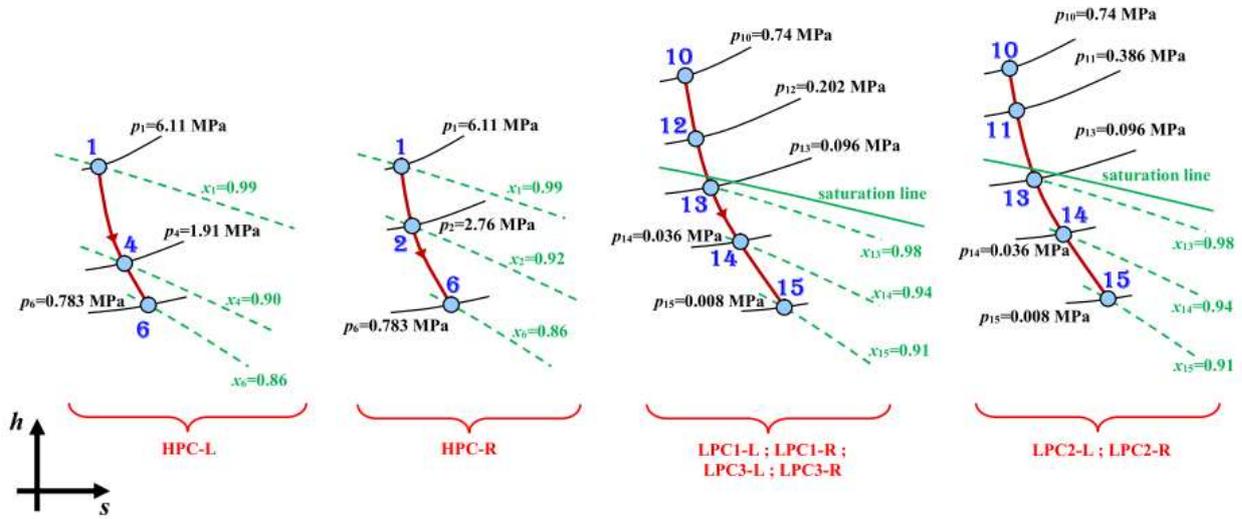

**Fig. 2** Real (polytropic) steam expansion processes in specific enthalpy – specific entropy (*h-s*) diagrams for each observed steam turbine part (Original operating regime)
**Source:** Authors

## 5 The influence of each observed optimization algorithm on the overall efficiency of the whole nuclear power plant

In this section is presented how the observed optimization algorithms (IGSA, GA and SA) influenced the overall efficiency of the entire power plant. By knowing that element, it can be predicted a behavior of certain turbine operating parameters related to the exergy analysis and it can be seen several relations between used optimization algorithms as well as the relation between the Original operating regime and obtained operating regime by each algorithm.

The overall efficiency of the entire nuclear power plant can be calculated by using an equation [104]:

$$\eta_{\text{overall}} = \frac{P_{\text{WT}}}{\dot{Q}_{\text{SG}}}, \tag{34}$$

where $\dot{Q}_{\text{SG}}$ is the amount of heat transferred to water in steam generators. $\dot{Q}_{\text{SG}}$ is directly proportional to the amount of heat delivered to HPC, which is calculated as [105]:

$$\dot{E}n_1 = \dot{m}_1 \cdot h_1, \tag{35}$$

where $\dot{E}n$ is a total steam energy flow, while index 1 is related to the HPC inlet (according to Fig. 1). The amount of heat delivered to HPC in all four observed operating regimes (Original, IGSA, GA and SA) is presented in Fig. 3. In comparison to the Original operating regime, all used optimization algorithms increases the amount of heat delivered to HPC (and consequentially amount of heat transferred to water in steam generators). Also, from Fig. 3 is clear that the highest amount of heat delivered to HPC is performed by using IGSA, followed by GA, while the lowest amount of heat delivered to HPC is performed by the SA (in comparison to other algorithms).

By following conclusions from [97] that the highest increase in overall power plant efficiency (in comparison to Original operating regime) is obtained by IGSA, followed by GA and finally SA and according to Eq. 34, it can be expected that the highest increase in developed mechanical power of the whole turbine (again, in comparison to Original operating regime) will be obtained by IGSA, followed by GA and finally by SA. However, this conclusion is related to increase in mechanical power of the whole turbine, from it cannot be seen how each algorithm influenced turbine cylinders and cylinder parts from the developed mechanical power viewpoint.

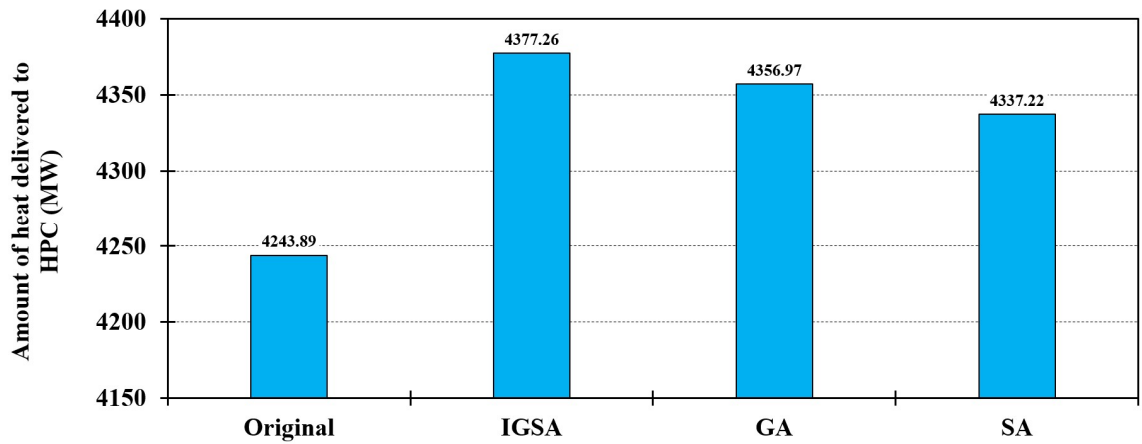

**Fig. 3** Amount of heat delivered to HPC – Original operating regime and operating regimes obtained by three optimization algorithms (IGSA, GA and SA)
**Source:** Authors

## 6 Exergy analysis of the observed steam turbine

The first element in exergy analysis of any steam turbine, its cylinders or cylinder parts is a calculation of developed mechanical power (according to real – polytropic steam expansion process, Fig. 2).

Mechanical power developed by the whole turbine in Original operating regime is equal to 996.29 MW and is notably lower in comparison to developed mechanical power of each observed algorithm, Fig. 4. Operating regime obtained by IGSA gives the highest developed mechanical power of the whole turbine equal to 1022.48 MW, followed by GA (1020.06 MW), while the lowest developed mechanical power of the whole turbine is obtained by SA (1017.16 MW) if observing all the algorithms, Fig. 4. However, even by using SA the developed mechanical power of the whole turbine is notably higher than in the Original operating regime. Obtained results confirm conclusion from a previous section that the highest developed mechanical power of the whole turbine will be obtained by IGSA, followed by GA and SA, while the lowest mechanical power of the whole turbine will be developed in Original operating regime.

From Fig. 3 and Fig. 4 can be concluded that each optimization algorithm increases the amount of heat transferred to water in steam generators and simultaneously increases developed mechanical power of the whole turbine. The increase ratio of the whole turbine developed mechanical power is higher than the increase ratio of transferred heat amount in steam generators, so by using optimization algorithms overall plant efficiency increases (Eq. 34).

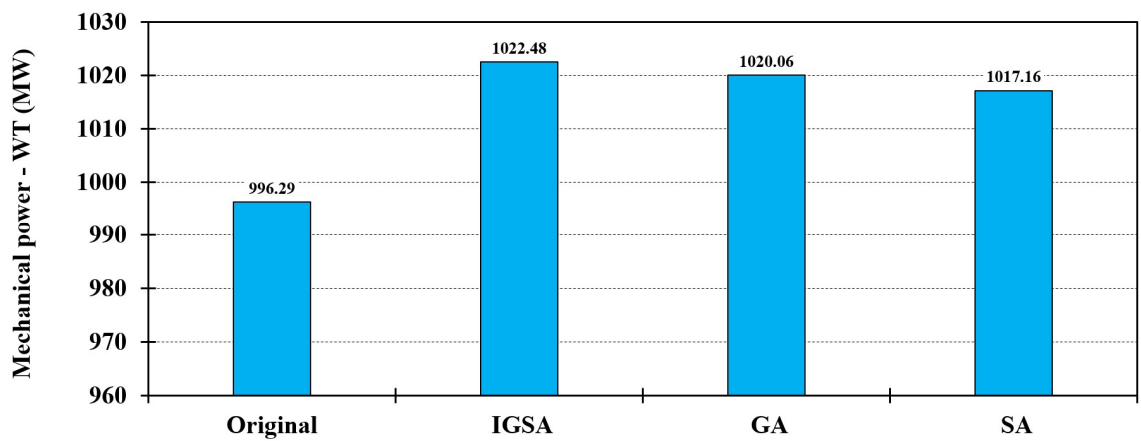

**Fig. 4** Mechanical power developed by the whole turbine (WT) – Original operating regime and operating regimes obtained by three optimization algorithms (IGSA, GA and SA)
**Source:** Authors

The influence of all optimization algorithms on the developed mechanical power of each cylinder and cylinder part can be clearly seen in Fig. 5. In general, all optimization algorithms (IGSA, GA and SA) increases developed mechanical power of each cylinder or cylinder part (in comparison to Original operating regime). However, its effect and increase is not the same for each cylinder.

While observing entire HPC (HPC-EC) it can be seen that the highest increase in developed mechanical power is obtained by using GA, followed by IGSA and SA. Therefore, by observing entire HPC only, Genetic Algorithm is more beneficial for the mechanical power increase than the other two algorithms. Also, by observing both HPC parts, it can be seen that for left HPC part (HPC-L) the highest mechanical power is developed by using IGSA, while observing right HPC part (HPC-R), the highest mechanical power is developed in GA operating regime. Therefore, for the entire HPC can be concluded that due to the cylinder asymmetry, different optimization algorithms have various influences of both cylinder parts (when the developed mechanical power is taken into consideration).

Observed optimization algorithms have the same influence on the developed mechanical power of all low pressure cylinders (LPC1, LPC2 and LPC3) and each of its parts, Fig. 5. In comparison to the Original operating regime, the IGSA result with the highest increase in mechanical power of each low pressure cylinder and its part, followed by GA and finally SA.

From Fig. 5 can be concluded that IGSA has a dominant influence on the developed mechanical power of the observed turbine because in comparison to Original operating regime, it dominantly increases mechanical power of those cylinders which cumulatively develop the highest part of mechanical power (low pressure cylinders). Regardless of the fact that increase in mechanical power of HPC is higher by using GA than IGSA, the HPC develops lower mechanical power in comparison to all low pressure cylinders.

From the above observations can be concluded that observed optimization algorithms did not have the same influence for all turbine cylinders (related to developed mechanical power). The dominant increase in mechanical power of the whole turbine (in comparison to Original operating regime) will have an optimization algorithm which dominantly increases the mechanical power of the most (not necessarily all) turbine cylinders. Also, from the viewpoint of developed mechanical power, both parts of dual flow asymmetrical cylinders can be variously influenced if several optimization algorithms are considered, while such occurrence cannot be expected in symmetrical dual flow cylinders.

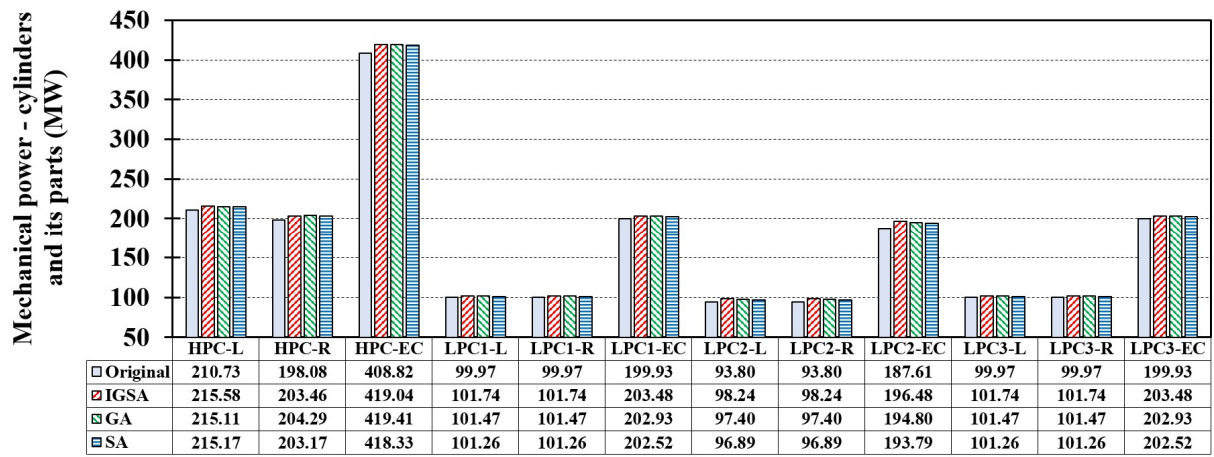

**Fig. 5** Mechanical power developed by the turbine cylinders and its parts – Original operating regime and operating regimes obtained by three optimization algorithms (IGSA, GA and SA)
**Source:** Authors

### 6.1. Exergy analysis at the base ambient state

At the base ambient state it is performed calculation of exergy destruction and exergy efficiency of the whole turbine, turbine cylinders and cylinder parts in Original operating regime as well as in operating regimes obtained by three optimization algorithms (IGSA, GA and SA).

Fig. 6 shows the exergy destruction of the whole turbine at the base ambient state in each four observed operating regimes. Previously, it is confirmed that each optimization algorithm increases the amount of heat delivered to HPC and developed mechanical power of the whole turbine (in comparison to Original operating regime), but from Fig. 6 it can also be seen that all observed optimization algorithms increases whole turbine exergy destruction.

IGSA which increases whole turbine developed mechanical power and the amount of heat delivered to HPC the most, simultaneously results with the lowest exergy destruction of the whole turbine (in comparison to other algorithms). Operating regime obtained by GA slightly increases whole turbine exergy destruction in comparison to IGSA (167.56 MW in comparison to 167.49 MW), while SA resulted with the notably higher exergy destruction of the whole turbine in comparison to GA and IGSA (equal to 168.07 MW).

From the viewpoint of the exergy destruction can be concluded that the lowest increase in the exergy destruction of the whole turbine (in comparison to Original operating regime) causes the optimization algorithm which increases developed mechanical power of the whole turbine more than others (in this case that is IGSA).

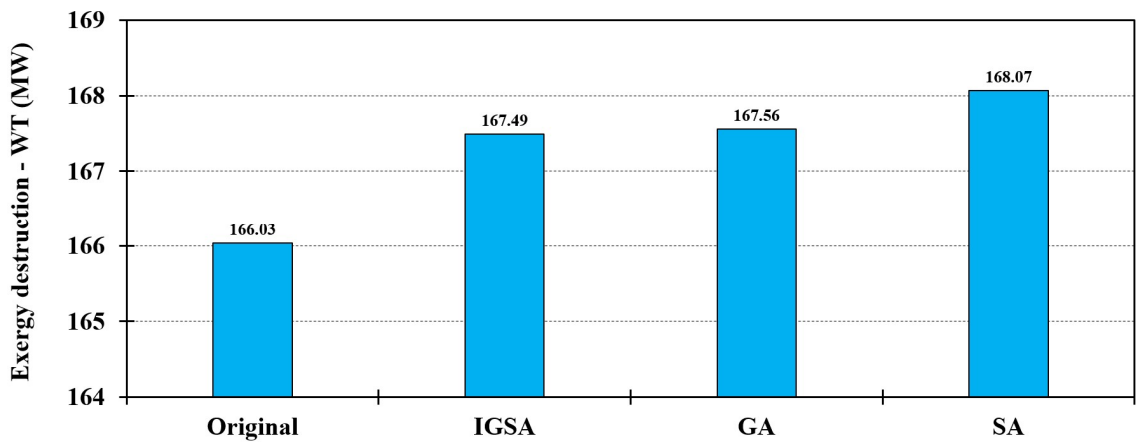

**Fig. 6** Exergy destruction of the whole turbine (WT) at the base ambient state – Original operating regime and operating regimes obtained by three optimization algorithms (IGSA, GA and SA)

**Source:** Authors

Change in the exergy destruction of turbine cylinders and its parts in Original operating regime and operating regimes obtained by three optimization algorithms (IGSA, GA and SA) is presented in Fig. 7.

For the entire HPC and both of its parts it can be stated that all optimization algorithms increases exergy destruction in comparison to the Original operating regime. For the left part of HPC (HPC-L) the lowest increase in exergy destruction in comparison to the Original operating regime is obtained by using GA, while simultaneously, for the right HPC part (HPC-R) GA causes the highest exergy destruction when compared to other algorithms. Exergy destruction of the entire HPC (HPC-EC) is increased the lowest (in comparison to Original operating regime) by using GA, following by IGSA and finally by SA, Fig. 7. As well as developed mechanical power, exergy destruction also proves that the left and right parts of asymmetrical cylinders can be influenced completely differently while using various optimization algorithms.

Similar as for entire HPC and both of its parts, all optimization algorithms increases the exergy destruction of entire LPC2 and both of its parts in comparison to the Original operating regime. The lowest increase in the exergy destruction of entire LPC2 and both of its parts (in comparison to Original operating regime) can be seen if the GA is used, followed by SA, while IGSA causes the highest exergy destruction of entire LPC2 and both of its parts.

All observed optimization algorithms decreases exergy destruction of entire LPC1 and LPC3 as well as both parts of each cylinder in comparison to the Original operating regime. By observing three analyzed algorithms, it can be concluded that the lowest exergy destruction of entire LPC1 and LPC3 as well as both of its parts is obtained by using IGSA, while the highest exergy destruction is obtained by using SA.

From the viewpoint of exergy destruction it can be concluded that GA has the most beneficial influence on entire HPC and LPC2, while IGSA results with the lowest exergy destruction of LPC1 and LPC3. However, parts of each cylinder can be differently influenced by various algorithms in terms of exergy destruction. Observing only the exergy destruction of entire turbine cylinders and its parts, it can be concluded that the optimal solution will be the usage of analyzed algorithms combination.

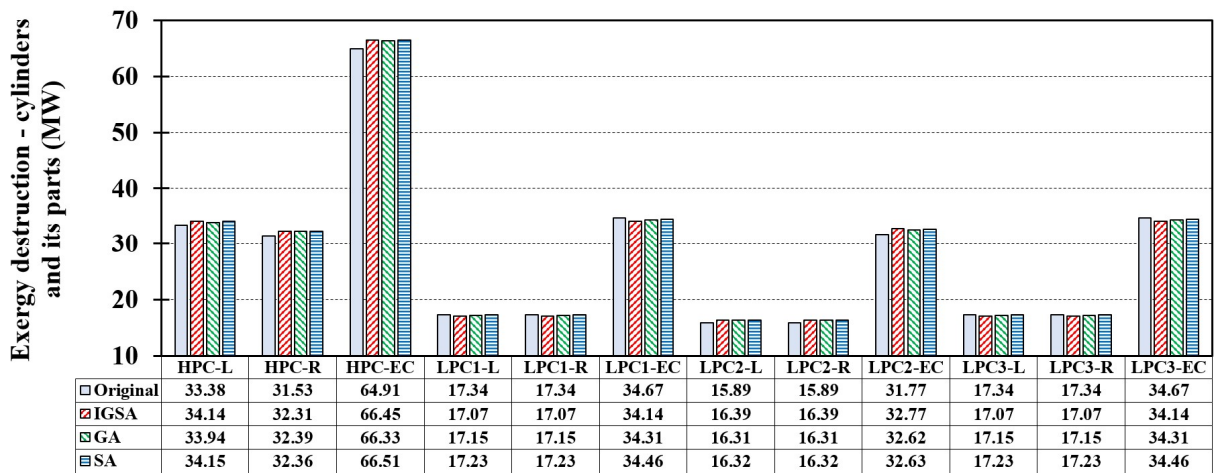

**Fig. 7** Exergy destruction of the turbine cylinders and its parts at the base ambient state –
Original operating regime and operating regimes obtained by three optimization algorithms
(IGSA, GA and SA)

**Source:** Authors

Exergy efficiencies of the whole turbine, turbine cylinders and its parts in all four observed operating regimes are presented in Fig. 8.

When observing entire HPC and both of its parts, it can be seen that GA and IGSA increases, while SA decreases exergy efficiencies in comparison to the Original operating regime. The highest exergy efficiencies of entire HPC and both of its parts are obtained by using GA (when considering all four operating regimes). In comparison to all low pressure cylinders and its parts, it can be stated that HPC and both of its parts have higher exergy efficiencies (around 1% higher).

All three optimization algorithms increases exergy efficiencies of all low pressure cylinders and its parts in comparison to the Original operating regime. The highest exergy efficiencies of all low pressure cylinders and its parts are obtained by using IGSA, followed by GA and finally SA. Also, from Fig. 8 can be noted that all optimization algorithms notably increases exergy efficiencies of LPC1 and LPC3 (and both its parts) in comparison to the Original operating regime, while the increase in exergy efficiency is still visible for LPC2 and both its parts, but not so notable.

The highest exergy efficiency of the whole turbine, equal to 85.92% is obtained by IGSA, followed by GA (85.89%) and SA (85.82%), while the lowest exergy efficiency of the whole turbine can be observed in Original operating regime (85.70%). Exergy efficiencies of the whole observed turbine, turbine cylinders and its parts are lower in comparison to steam turbines and its cylinders which dominantly operates by using superheated steam [76, 104].

Final conclusion which can be derived from Fig. 8 is that IGSA causes the highest increase in exergy efficiencies of the whole turbine and all low pressure turbine cylinders (in comparison to the Original operating regime and other algorithms). Only for the entire HPC and both its parts, GA shows dominancy related to exergy efficiencies in comparison to all other operating regimes.

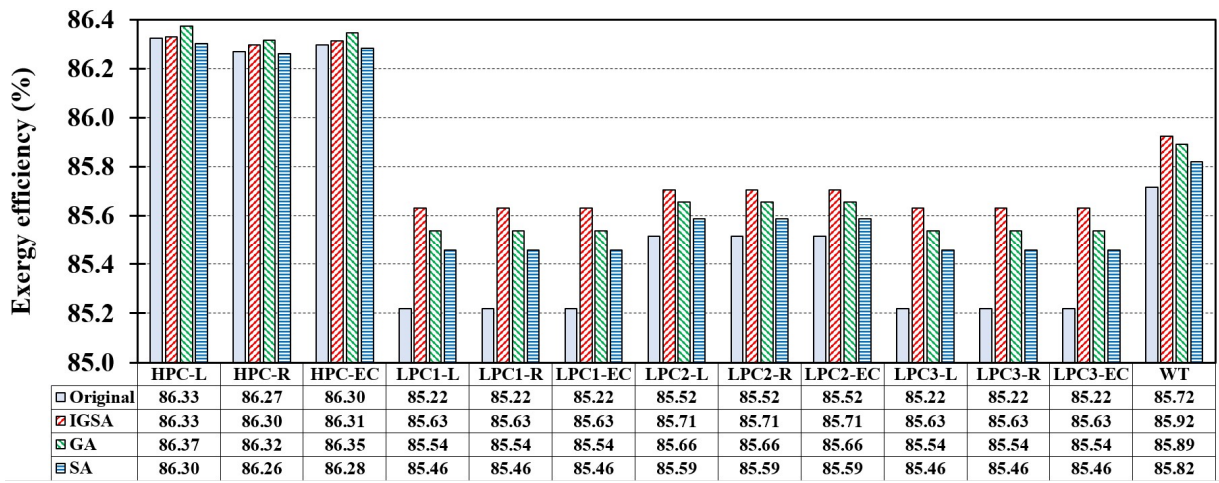

**Fig. 8** Exergy efficiency of the whole turbine, turbine cylinders and its parts at the base ambient state – Original operating regime and operating regimes obtained by three optimization algorithms (IGSA, GA and SA)

**Source:** Authors

Finally, it should be noted that the same relations about exergy destructions and exergy efficiencies of the whole turbine, turbine cylinders and its parts at the base ambient state are obtained at all the other observed ambient temperatures. Therefore, for the observed steam turbine it can be stated that obtained relations are valid in general, regardless of the selected ambient state.

## 6.2 Exergy analysis during the ambient temperature change

The average change in exergy destruction between the ambient temperatures 5 °C and 45 °C (in steps of 10 °C) of the whole turbine, turbine cylinders and its parts is presented in Fig. 9 in all four observed operating regimes.

For the entire HPC can be clearly seen that exergy destruction change will be the highest in operating regime obtained by IGSA, while the lowest will be in the Original operating regime. Both HPC parts again differs in the sensitivity of exergy destruction change related to the ambient temperature change. HPC-L will have the highest exergy destruction change by using IGSA, while HPC-R will have the highest exergy destruction change by using SA. For the entire HPC and both its parts is valid that exergy destruction will be more sensitive by the ambient temperature change in operating regimes obtained with optimization algorithms than in the Original operating regime.

Exergy destruction of entire LPC2 and both its parts will be more sensitive by the ambient temperature change in operating regimes obtained with optimization algorithms than in the Original operating regime. Exergy destruction change of LPC2 and both its parts will be the highest in IGSA operating regime.

Exergy destruction of entire LPC1 and LPC3 as well as both its parts is the highest influenced by the ambient temperature change in the Original operating regime, Fig. 9. The lowest sensitivity to the ambient temperature change, related to the exergy destruction of entire LPC1 and LPC3 as well as both its parts shows IGSA, followed by GA and SA.

Exergy destruction of the whole turbine will be more sensitive by the ambient temperature change in operating regimes obtained with optimization algorithms than in the Original operating regime. Observing operating regimes which involved all three optimization algorithms, it can be concluded that the exergy destruction of the whole turbine shows the highest sensitivity related to the ambient temperature change in SA, followed by IGSA and GA operating regimes.

General conclusion related to the average change in exergy destruction during the ambient temperature change is that the whole turbine, turbine cylinders and its parts are more influenced in operating regimes obtained by three optimization algorithms than in the Original operating regime. Also, exergy destruction of HPC and both its parts is much more influenced by the ambient temperature change in comparison to all low pressure cylinders and its parts.

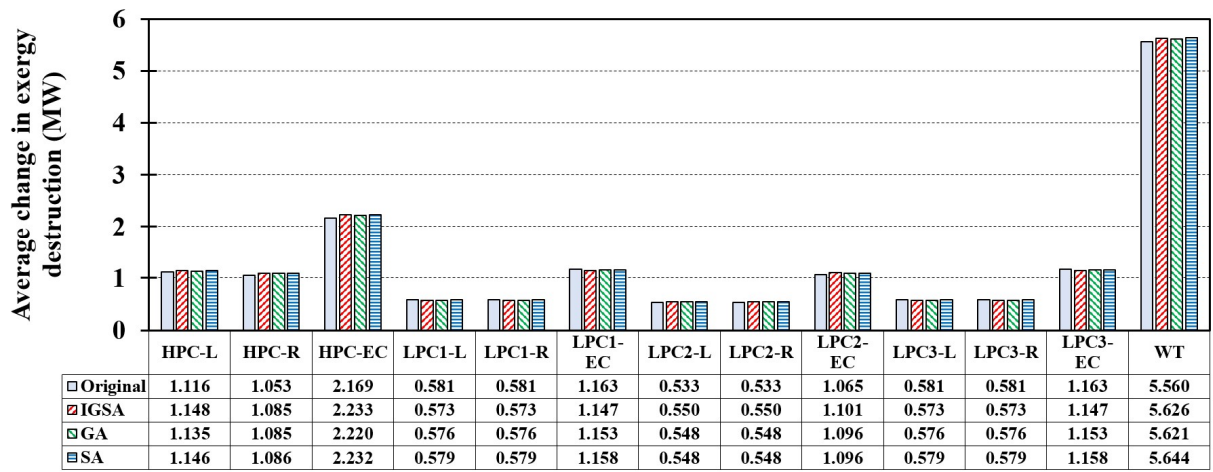

**Fig. 9** Average change in exergy destruction between the ambient temperatures 5 °C and 45 °C (in steps of 10 °C) of the whole turbine, turbine cylinders and its parts – Original operating regime and operating regimes obtained by three optimization algorithms (IGSA, GA and SA)
**Source:** Authors

The average change in exergy efficiency between the ambient temperatures 5 °C and 45 °C (in steps of 10 °C) of the whole turbine, turbine cylinders and its parts in all four observed operating regimes is presented in Fig. 10.

From Fig. 10 can be clearly seen that exergy efficiency of HPC and both its parts is much lower influenced by the ambient temperature change in comparison to all low pressure cylinders and its parts. The exergy efficiency of the entire HPC and both its parts is much lower influenced by the ambient temperature change in Original and GA operating regimes in comparison to IGSA and SA operating regimes.

The exergy efficiency of all low pressure cylinders and its parts is the highest influenced by the ambient temperature change in the Original operating regime in comparison to any other operating regime obtained with optimization algorithms. Also, from Fig. 10 can clearly be seen that exergy efficiency of all low pressure cylinders and its parts is the lowest influenced by the ambient temperature change in IGSA operating regime (followed by GA and SA operating regimes).

The exergy efficiency of the whole turbine is highest influenced by the ambient temperature change in the Original operating regime, followed by the SA operating regime. IGSA and GA operating regimes show the lowest sensitivity related to exergy efficiency during the ambient temperature change.

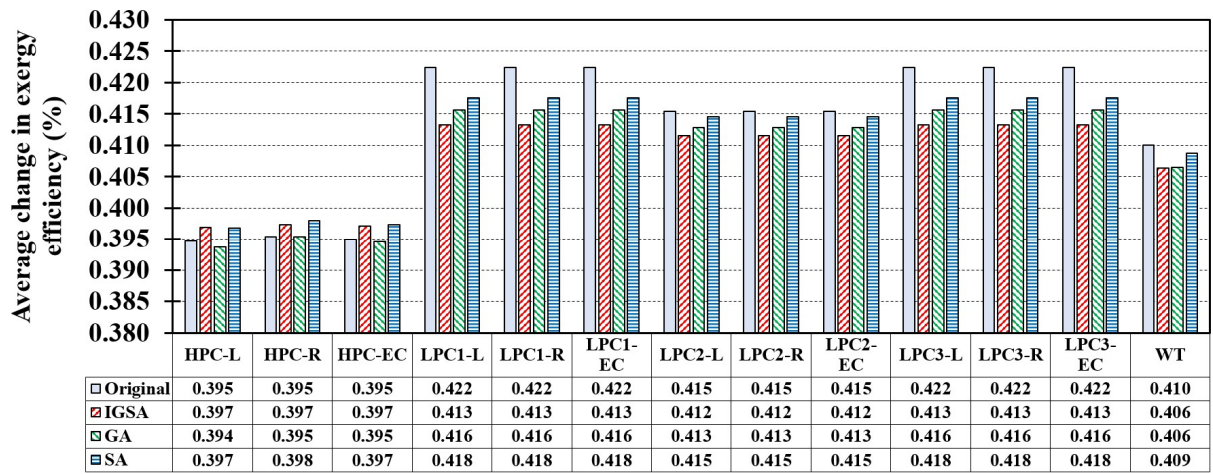

**Fig. 10** Average change in exergy efficiency between the ambient temperatures 5 °C and 45 °C (in steps of 10 °C) of the whole turbine, turbine cylinders and its parts – Original operating regime and operating regimes obtained by three optimization algorithms (IGSA, GA and SA)
**Source:** Authors

Exergy analysis of the whole turbine, turbine cylinders and its parts during the ambient temperature change shows that this turbine (which dominantly operate by using wet steam) is low influenced by the ambient temperature change. This fact is confirmed from Fig. 10 because the ambient temperature change of 10 °C results with an average change in exergy efficiency lower than 0.5% for the whole turbine, turbine cylinders and each cylinder part. The same conclusion can be found in the literature [77] for a steam turbines which dominantly operate by using superheated steam. The exergy efficiency of the analyzed steam turbine is low sensitive by the ambient temperature change what significantly differs in comparison to some other components (for example, low pressure condensate heaters) which also uses wet steam for its operation [98].

Further research and analysis of the whole observed steam turbine, turbine cylinders and cylinder parts will be performed by using other optimization algorithms and various artificial intelligence methods which proved its applicability and good performance not only in power systems, but also in other engineering fields [106-109]. It will be investigated can some other optimization algorithm or method show better performance related to this turbine exergy destruction and efficiency.

## 7 Conclusions

In this paper is performed exergy analysis of the whole steam turbine from nuclear power plant, turbine cylinders and each cylinder part in four different operating regimes: Original operating regime (standard turbine operation) and three operating regimes obtained by different optimization algorithms. Optimization algorithms: SA (Simplex Algorithm), GA (Genetic Algorithm) and IGSA (Improved Genetic-Simplex Algorithm) are basically used for the overall nuclear power plant efficiency increase. However, from the literature is not known how each of these algorithms influenced efficiencies and destructions of the whole steam turbine, turbine cylinders and each cylinder part. In addition, in this research is investigated did the ambient temperature change influenced steam turbine from nuclear power plant (which dominantly operates by using wet steam) more than turbine which dominantly operate by using superheated steam (turbine from conventional or marine steam power plant). The most important conclusions obtained in this research are:
- All observed optimization algorithms increases overall plant efficiency (in comparison to the Original operating regime) by the simultaneous increase of heat amount transferred to water in steam generators and increase of whole turbine developed mechanical power. In this relation, increase in mechanical power has a higher intensity than the increase in heat amount transferred to water.
- Whole turbine develop higher mechanical power in any operating regime obtained by optimization algorithms in comparison to the Original operating regime. Operating regime obtained by IGSA gives the highest developed mechanical power of the whole turbine equal to 1022.48 MW, followed by GA (1020.06 MW) and SA (1017.16 MW), while in Original operating regime whole turbine develop mechanical power equal to 996.29 MW.
- IGSA causes the highest increase in developed mechanical power of almost all cylinders and cylinder parts (in relation to the Original operating regime). Only the entire HPC and its right part (HPC-R) shows better performance in the developed mechanical power if the GA is used (instead of IGSA).
- All observed optimization algorithms increases the exergy destruction of the whole turbine (in comparison to the Original operating regime). The lowest increase in the exergy destruction of the whole turbine (in comparison to the Original operating regime) causes IGSA, followed by GA, while SA causes the highest increase in the exergy destruction of the whole turbine.
- GA has the most beneficial influence on the exergy destruction of entire HPC and LPC2, while IGSA results with the lowest exergy destruction of LPC1 and LPC3. However, parts of each cylinder can be differently influenced by various algorithms in terms of exergy destruction. The optimal solution in terms of exergy destruction decrease of entire turbine cylinders and its parts will be the usage of analyzed algorithms combination.
- When considering all four operating regimes, the highest exergy efficiencies of entire HPC and both of its parts are obtained by using GA. The highest exergy efficiencies of all low pressure cylinders and its parts are obtained by using IGSA. IGSA also shows its dominancy in terms of exergy efficiency if the whole turbine is observed - the highest exergy efficiency of the whole turbine, equal to 85.92% is obtained by IGSA, followed by GA (85.89%) and SA (85.82%), while the lowest exergy efficiency of the whole turbine is obtained in the Original operating regime (85.7%).
- Analyzed turbine, which dominantly operates by using wet steam is low influenced by the ambient temperature change because the ambient temperature change of 10 °C resulted with an average change in exergy efficiency lower than 0.5% for the whole turbine, turbine cylinders and each cylinder part. Therefore, from the viewpoint of the ambient temperature change – there is no notable difference between steam turbines which dominantly operate by using wet or superheated steam.

**Acknowledgment**

This research has been supported by the Croatian Science Foundation under the project IP-2018-01-3739, CEEPUS network CIII-HR-0108, European Regional Development Fund under the grant KK.01.1.1.01.0009 (DATACROSS), project CEKOM under the grant KK.01.2.2.03.0004, CEI project "COVIDAi" (305.6019-20), University of Rijeka scientific grants: uniri-tehnic-18-275-1447, uniri-tehnic-18-18-1146 and uniri-tehnic-18-14.

**References**

[1] Tanuma, T. (Ed.). (2017). Advances in Steam Turbines for Modern Power Plants. Woodhead Publishing.
[2] Sutton, I. (2017). Plant design and operations. Gulf Professional Publishing.
[3] Elhelw, M., Al Dahma, K. S., & el Hamid Attia, A. (2019). Utilizing exergy analysis in studying the performance of steam power plant at two different operation mode. Applied Thermal Engineering, 150, 285-293. (doi:10.1016/j.applthermaleng.2019.01.003)

[4]  Kumar, V., Pandya, B., & Matawala, V. (2019). Thermodynamic studies and parametric effects on exergetic performance of a steam power plant. International Journal of Ambient Energy, 40(1), 1-11. (doi:10.1080/01430750.2017.1354326)

[5]  Naserbegi, A., Aghaie, M., Minuchehr, A., & Alahyarizadeh, G. (2018). A novel exergy optimization of Bushehr nuclear power plant by gravitational search algorithm (GSA). Energy, 148, 373-385. (doi:10.1016/j.energy.2018.01.119)

[6]  Marques, J. G., Costa, A. L., Pereira, C., & Fortini, Â. (2019). Energy and Exergy Analyses of Angra 2 Nuclear Power Plant. Brazilian Journal of Radiation Sciences, 7(2B). (doi:10.15392/bjrs.v7i2B.647)

[7]  Ahmad, A. D., Abubaker, A. M., Najjar, Y. S., & Manaserh, Y. M. A. (2020). Power boosting of a combined cycle power plant in Jordan: An integration of hybrid inlet cooling & solar systems. Energy Conversion and Management, 214, 112894. (doi:10.1016/j.enconman.2020.112894)

[8]  Lorencin, I., Anđelić, N., Mrzljak, V., & Car, Z. (2019). Genetic Algorithm Approach to Design of Multi-Layer Perceptron for Combined Cycle Power Plant Electrical Power Output Estimation. Energies, 12(22), 4352. (doi:10.3390/en12224352)

[9]  Burin, E. K., Vogel, T., Multhaupt, S., Thelen, A., Oeljeklaus, G., Görner, K., & Bazzo, E. (2016). Thermodynamic and economic evaluation of a solar aided sugarcane bagasse cogeneration power plant. Energy, 117, 416-428. (doi:10.1016/j.energy.2016.06.071)

[10] González Gutiérrez, C., Suárez de la Fuente, S., Bonello, J., & Bucknall, R. (2020). An empirical analysis on the operational profile of liquefied natural gas carriers with steam propulsion plants. Journal of Navigation, 1-20. (doi:10.1017/S0373463320000612)

[11] Koroglu, T., & Sogut, O. S. (2018). Conventional and advanced exergy analyses of a marine steam power plant. Energy, 163, 392-403. (doi:10.1016/j.energy.2018.08.119)

[12] Mrzljak, V., Poljak, I., & Prpić-Oršić, J. (2019). Exergy analysis of the main propulsion steam turbine from marine propulsion plant. Brodogradnja: Teorija i praksa brodogradnje i pomorske tehnike, 70(1), 59-77. (doi:10.21278/brod70105)

[13] Taheri, M. H., Mosaffa, A. H., & Farshi, L. G. (2017). Energy, exergy and economic assessments of a novel integrated biomass based multigeneration energy system with hydrogen production and LNG regasification cycle. Energy, 125, 162-177. (doi:10.1016/j.energy.2017.02.124)

[14] Alirahmi, S. M., Dabbagh, S. R., Ahmadi, P., & Wongwises, S. (2020). Multi-objective design optimization of a multi-generation energy system based on geothermal and solar energy. Energy Conversion and Management, 205, 112426. (doi:10.1016/j.enconman.2019.112426)

[15] Variny, M., & Janošovský, J. (2020). Reevaluation of optimal turbine/motor process drivers selection concept proposed by Wu et al. (2016). Journal of Cleaner Production, 253, 120029. (doi:10.1016/j.jclepro.2020.120029)

[16] Hafdhi, F., Khir, T., Yahyia, A. B., & Brahim, A. B. (2015). Energetic and exergetic analysis of a steam turbine power plant in an existing phosphoric acid factory. Energy Conversion and Management, 106, 1230-1241. (doi:10.1016/j.enconman.2015.10.044)

[17] Mrzljak, V., Poljak, I., & Mrakovčić, T. (2017). Energy and exergy analysis of the turbo-generators and steam turbine for the main feed water pump drive on LNG carrier. Energy conversion and management, 140, 307-323. (doi:10.1016/j.enconman.2017.03.007)

[18] Mrzljak, V. (2018). Low power steam turbine energy efficiency and losses during the developed power variation. Tehnički glasnik, 12(3), 174-180. (doi:10.31803/tg-20180201002943)

[19] Fernández, I. A., Gómez, M. R., Gómez, J. R., & Insua, Á. B. (2017). Review of propulsion systems on LNG carriers. Renewable and Sustainable Energy Reviews, 67, 1395-1411. (doi:10.1016/j.rser.2016.09.095)

[20] Mrzljak, V., & Poljak, I. (2019). Energy analysis of main propulsion steam turbine from conventional LNG carrier at three different loads. NAŠE MORE: znanstveni časopis za more i pomorstvo, 66(1), 10-18. (doi:10.17818/NM/2019/1.2)

[21] Behrendt, C., & Stoyanov, R. (2018). Operational Characteristic of Selected Marine Turbounits Powered by Steam from Auxiliary Oil-Fired Boilers. New Trends in Production Engineering, 1(1), 495-501. (doi:10.2478/ntpe-2018-0061)

[22] Elčić, Z. (1995). Steam turbines. ABB, Karlovac, National and University Library Zagreb.

[23] Noroozian, A., Mohammadi, A., Bidi, M., & Ahmadi, M. H. (2017). Energy, exergy and economic analyses of a novel system to recover waste heat and water in steam power plants. Energy conversion and management, 144, 351-360. (doi:10.1016/j.enconman.2017.04.067)

[24] Sadeghi, K., Ghazaie, S. H., Sokolova, E., Fedorovich, E., & Shirani, A. (2020). Comprehensive techno-economic analysis of integrated nuclear power plant equipped with various hybrid desalination systems. Desalination, 493, 114623. (doi:10.1016/j.desal.2020.114623)

[25] Kocijel, L., Poljak, I., Mrzljak, V., & Car, Z. (2020). Energy Loss Analysis at the Gland Seals of a Marine Turbo-Generator Steam Turbine. Tehnički glasnik, 14(1), 19-26. (doi:10.31803/tg-20191031094436)

[26] Zhao, Z., Su, S., Si, N., Hu, S., Wang, Y., Xu, J., ... & Xiang, J. (2017). Exergy analysis of the turbine system in a 1000 MW double reheat ultra-supercritical power plant. Energy, 119, 540-548. (doi:10.1016/j.energy.2016.12.072)

[27] Adibhatla, S., & Kaushik, S. C. (2014). Energy and exergy analysis of a super critical thermal power plant at various load conditions under constant and pure sliding pressure operation. Applied thermal engineering, 73(1), 51-65. (doi:10.1016/j.applthermaleng.2014.07.030)

[28] Selimli, S., & Sunay, S. (2020). Feasibility study of the energy and economic gain that can be achieved by driving the boiler feedwater pump with a backpressure steam turbine. Proceedings of the Institution of Mechanical Engineers, Part A: Journal of Power and Energy, 0957650920969466. (doi:10.1177/0957650920969466)

[29] Medica-Viola, V., Baressi Šegota, S., Mrzljak, V., & Štifanić, D. (2020). Comparison of conventional and heat balance based energy analyses of steam turbine. Pomorstvo, 34(1), 74-85. (doi:10.31217/p.34.1.9)

[30] Kozman, T. A., Liu, Y. C., Simon, W. E., & Guidry, J. J. (2012). A Revitalized Steam Lab and Its Impacts on Both Industry and Engineering Education. International Journal of Energy Engineering, 2(4), 114-118. (doi:10.5923/j.ijee.20120204.01)

[31] Nikam, K. C., Kumar, R., & Jilte, R. (2020). Economic and exergoeconomic investigation of 660 MW coal-fired power plant. Journal of Thermal Analysis and Calorimetry, 1-15. (doi:10.1007/s10973-020-10213-z)

[32] Kostyuk, A., Frolov, V.: Steam and gas turbines, Mir Publishers, Moscow, 1988.

[33] Hanus, K., Variny, M., & Illés, P. (2020). Assessment and Prediction of Complex Industrial Steam Network Operation by Combined Thermo-Hydrodynamic Modeling. Processes, 8(5), 622. (doi:10.3390/pr8050622)

[34] Opriş, I., Cenuşă, V. E., Norişor, M., Darie, G., Alexe, F. N., & Costinaş, S. (2020). Parametric optimization of the thermodynamic cycle design for supercritical steam power plants. Energy Conversion and Management, 208, 112587. (doi:10.1016/j.enconman.2020.112587)

[35] Medica-Viola, V., Mrzljak, V., Anđelić, N., & Jelić, M. (2020). Analysis of Low-Power Steam Turbine With One Extraction for Marine Applications. NAŠE MORE: znanstveni časopis za more i pomorstvo, 67(2), 87-95. (doi:10.17818/NM/2020/2.1)

[36] Blažević, S., Mrzljak, V., Anđelić, N., & Car, Z. (2019). Comparison of energy flow stream and isentropic method for steam turbine energy analysis. Acta Polytechnica, 59(2), 109-125. (doi:10.14311/AP.2019.59.0109)

[37] Zueco, J., López-Asensio, D., Fernández, F. J., & López-González, L. M. (2020). Exergy analysis of a steam-turbine power plant using thermocombustion. Applied Thermal Engineering, 180, 115812. (doi:10.1016/j.applthermaleng.2020.115812)

[38] Mrzljak, V., Senčić, T., & Žarković, B. (2018). Turbogenerator steam turbine variation in developed power: Analysis of exergy efficiency and exergy destruction change. Modelling and Simulation in Engineering, 2018. (doi:10.1155/2018/2945325)

[39] Kumar, A. K. B. A., Nikam, K. C., & Behura, A. K. (2020). An exergy analysis of a 250 MW thermal power plant. Renewable Energy Research and Application, 1(2), 197-204. (doi:10.22044/rera.2020.9460.1025)

[40] Uysal, C., Kurt, H., & Kwak, H. Y. (2017). Exergetic and thermoeconomic analyses of a coal-fired power plant. International Journal of Thermal Sciences, 117, 106-120. (doi:10.1016/j.ijthermalsci.2017.03.010)

[41] Ogorure, O. J., Oko, C. O. C., Diemuodeke, E. O., & Owebor, K. (2018). Energy, exergy, environmental and economic analysis of an agricultural waste-to-energy integrated multigeneration thermal power plant. Energy conversion and management, 171, 222-240. (doi:10.1016/j.enconman.2018.05.093)

[42] Kowalczyk, T., Badur, J., & Bryk, M. (2019). Energy and exergy analysis of hydrogen production combined with electric energy generation in a nuclear cogeneration cycle. Energy Conversion and Management, 198, 111805. (doi:10.1016/j.enconman.2019.111805)

[43] Mrzljak, V., Poljak, I., & Medica-Viola, V. (2017). Dual fuel consumption and efficiency of marine steam generators for the propulsion of LNG carrier. Applied Thermal Engineering, 119, 331-346. (doi:10.1016/j.applthermaleng.2017.03.078)

[44] Hashemian, A., Lakzian, E., & Ebrahimi-Fizik, A. (2020). On the application of isogeometric finite volume method in numerical analysis of wet-steam flow through turbine cascades. Computers & Mathematics with Applications, 79(6), 1687-1705. (doi:10.1016/j.camwa.2019.09.025)

[45] Rzadkowski, R., Kubitz, L., Maziarz, M., Troka, P., Dominiczak, K., & Szczepanik, R. (2020). Tip-Timing measurements and numerical analysis of last-stage steam turbine mistuned bladed disc during run-down. Journal of Vibration Engineering & Technologies, 8(3), 409-415. (doi:10.1007/s42417-019-00185-2)

[46] Ebrahimgol, H., Aghaie, M., Zolfaghari, A., & Naserbegi, A. (2020). A novel approach in exergy optimization of a WWER1000 nuclear power plant using whale optimization algorithm. Annals of Nuclear Energy, 145, 107540. (doi:10.1016/j.anucene.2020.107540)

[47] Talebi, S., & Norouzi, N. (2020). Entropy and exergy analysis and optimization of the VVER nuclear power plant with a capacity of 1000 MW using the firefly optimization algorithm. Nuclear Engineering and Technology, 52(12), 2928-2938. (doi:10.1016/j.net.2020.05.011)

[48] Liu, M., Wang, S., & Yan, J. (2021). Operation scheduling of a coal-fired CHP station integrated with power-to-heat devices with detail CHP unit models by particle swarm optimization algorithm. Energy, 214, 119022. (doi:10.1016/j.energy.2020.119022)

[49] Sundaram, A. (2020). Multiobjective multi-verse optimization algorithm to solve combined economic, heat and power emission dispatch problems. Applied Soft Computing, 91, 106195. (doi:10.1016/j.asoc.2020.106195)

[50] Nuerlan, A., Wang, P., & Zhao, F. (2020). A neural network based inverse system control strategy to decouple turbine power in multi-reactor and multi-turbine nuclear power plant. Progress in Nuclear Energy, 129, 103500. (doi:10.1016/j.pnucene.2020.103500)

[51] Baressi Šegota, S., Lorencin, I., Anđelić, N., Mrzljak, V., & Car, Z. (2020). Improvement of Marine Steam Turbine Conventional Exergy Analysis by Neural Network Application. Journal of Marine Science and Engineering, 8(11), 884. (doi:10.3390/jmse8110884)

[52] Kim, M. J., Kim, T. S., Flores, R. J., & Brouwer, J. (2020). Neural-network-based optimization for economic dispatch of combined heat and power systems. Applied Energy, 265, 114785. (doi:10.1016/j.apenergy.2020.114785)

[53] Kumar, R., Nikam, K., & Jilte, R. (2020). A simulation model to predict coal-fired power plant production rate using artificial neural network tool. In Applied Computer Vision and Image Processing (pp. 150-160). Springer, Singapore. (doi:10.1007/978-981-15-4029-5_15)

[54] Durmayaz, A., & Yavuz, H. (2001). Exergy analysis of a pressurized-water reactor nuclear-power plant. Applied energy, 69(1), 39-57. (doi:10.1016/S0306-2619(00)00071-4)

[55] Terzi, R., Tükenmez, İ., & Kurt, E. (2016). Energy and exergy analyses of a VVER type nuclear power plant. International journal of hydrogen energy, 41(29), 12465-12476. (doi:10.1016/j.ijhydene.2016.02.016)

[56] Mrzljak, V., Poljak, I., & Medica-Viola, V. (2017). Thermodynamical analysis of high-pressure feed water heater in steam propulsion system during exploitation. Brodogradnja: Teorija i praksa brodogradnje i pomorske tehnike, 68(2), 45-61. (doi:10.21278/brod68204)

[57] Burin, E. K., Buranello, L., Giudice, P. L., Vogel, T., Görner, K., & Bazzo, E. (2015). Boosting power output of a sugarcane bagasse cogeneration plant using parabolic trough collectors in a feedwater heating scheme. Applied Energy, 154, 232-241. (doi:10.1016/j.apenergy.2015.04.100)

[58] Shamarokov, A. S., Zorin, V. M., & Bosenko, S. V. (2020). Optimization of Minimum Temperature Differences in the Steam Reheater of a Turbine at a Nuclear Power Station. Thermal Engineering, 67(3), 178-184. (doi:10.1134/S0040601520030052)

[59] Zhang, D., Yu, J., Tian, W., Su, G. H., & Qiu, S. (2019). Heat transfer characteristics in super-low finned-tube bundles of moisture separator reheaters. Nuclear Engineering and Design, 341, 368-376. (doi:10.1016/j.nucengdes.2018.11.018)

[60] Nandini, M., Sekhar, Y. R., & Subramanyam, G. (2021). Energy analysis and water conservation measures by water audit at thermal power stations. Sustainable Water Resources Management, 7(1), 1-24. (doi:10.1007/s40899-020-00487-4)

[61] Mrzljak, V., Poljak, I., & Medica-Viola, V. (2016). Efficiency and losses analysis of low-pressure feed water heater in steam propulsion system during ship maneuvering period. Pomorstvo, 30(2), 133-140. (doi:10.31217/p.30.2.6)

[62] Jamali, J. A. (2021). Steam condenser exergy analysis of steam power plant at different loads. North American Academic Research, 4(3), 21-31. (doi:10.5281/zenodo.4590310)

[63] Škopac, L., Medica-Viola, V., & Mrzljak, V. (2020). Selection Maps of Explicit Colebrook Approximations according to Calculation Time and Precision. Heat Transfer Engineering, 1-15. (doi:10.1080/01457632.2020.1744248)

[64] Kanoğlu, M., Çengel, Y. A., & Dinçer, İ. (2012). Efficiency evaluation of energy systems. Springer Science & Business Media.

[65] Dincer, I., & Rosen, M. A. (2012). Exergy: energy, environment and sustainable development. Newnes.

[66] Vundela Siva, R., Subash Chndra, K., Sudhir Kumar, T., & Narayanlal, P. (2010). An approach to analyse energy and exergy analysis of thermal power plants: a review. Smart Grid and Renewable Energy, 2010. (doi:10.4236/sgre.2010.13019)

[67] Poljak, I., Orović, J., Mrzljak, V., & Bernečić, D. (2020). Energy and exergy evaluation of a two-stage axial vapour compressor on the LNG carrier. Entropy, 22(1), 115. (doi:10.3390/e22010115)

[68] Baldi, F., Ahlgren, F., Nguyen, T. V., Thern, M., & Andersson, K. (2018). Energy and exergy analysis of a cruise ship. Energies, 11(10), 2508. (doi:10.3390/en11102508)

[69] Presciutti, A., Asdrubali, F., Baldinelli, G., Rotili, A., Malavasi, M., & Di Salvia, G. (2018). Energy and exergy analysis of glycerol combustion in an innovative flameless power plant. Journal of Cleaner Production, 172, 3817-3824. (doi:10.1016/j.jclepro.2017.06.022)

[70] Mrzljak, V., Prpić-Oršić, J., & Senčić, T. (2018). Change in steam generators main and auxiliary energy flow streams during the load increase of LNG carrier steam propulsion system. Pomorstvo, 32(1), 121-131. (doi:10.31217/p.32.1.15)

[71] Si, N., Zhao, Z., Su, S., Han, P., Sun, Z., Xu, J., ... & Xiang, J. (2017). Exergy analysis of a 1000 MW double reheat ultra-supercritical power plant. Energy Conversion and Management, 147, 155-165. (doi:10.1016/j.enconman.2017.05.045)

[72] Topal, H., Taner, T., Naqvi, S. A. H., Altınsoy, Y., Amirabedin, E., & Ozkaymak, M. (2017). Exergy analysis of a circulating fluidized bed power plant co-firing with olive pits: A case study of power plant in Turkey. Energy, 140, 40-46. (doi:10.1016/j.energy.2017.08.042)

[73] Vakilabadi, M. A., Bidi, M., Najafi, A. F., & Ahmadi, M. H. (2019). Exergy analysis of a hybrid solar-fossil fuel power plant. Energy Science & Engineering, 7(1), 146-161. (doi:10.1002/ese3.265)

[74] Ahmadi, G., Toghraie, D., & Akbari, O. (2019). Energy, exergy and environmental (3E) analysis of the existing CHP system in a petrochemical plant. Renewable and Sustainable Energy Reviews, 99, 234-242. (doi:10.1016/j.rser.2018.10.009)

[75] Aljundi, I. H. (2009). Energy and exergy analysis of a steam power plant in Jordan. Applied thermal engineering, 29(2-3), 324-328. (doi:10.1016/j.applthermaleng.2008.02.029)

[76] Ahmadi, G. R., & Toghraie, D. (2016). Energy and exergy analysis of Montazeri steam power plant in Iran. Renewable and Sustainable Energy Reviews, 56, 454-463. (doi:10.1016/j.rser.2015.11.074)

[77] Kopac, M., & Hilalci, A. (2007). Effect of ambient temperature on the efficiency of the regenerative and reheat Çatalağzı power plant in Turkey. Applied Thermal Engineering, 27(8-9), 1377-1385. (doi:10.1016/j.applthermaleng.2006.10.029)

[78] Mrzljak, V., Blecich, P., Anđelić, N., & Lorencin, I. (2019). Energy and Exergy Analyses of Forced Draft Fan for Marine Steam Propulsion System during Load Change. Journal of Marine Science and Engineering, 7(11), 381. (doi:10.3390/jmse7110381)

[79] Şöhret, Y., Açıkkalp, E., Hepbasli, A., & Karakoc, T. H. (2015). Advanced exergy analysis of an aircraft gas turbine engine: splitting exergy destructions into parts. Energy, 90, 1219-1228. (doi:10.1016/j.energy.2015.06.071)

[80] Ray, T. K., Datta, A., Gupta, A., & Ganguly, R. (2010). Exergy-based performance analysis for proper O&M decisions in a steam power plant. Energy Conversion and Management, 51(6), 1333-1344. (doi:10.1016/j.enconman.2010.01.012)

[81] Mrzljak, V., Poljak, I., & Medica-Viola, V. (2017). Energy and exergy efficiency analysis of sealing steam condenser in propulsion system of LNG carrier. NAŠE MORE: znanstveni časopis za more i pomorstvo, 64(1), 20-25. (doi:10.17818/NM/2017/1.4)

[82] Tan, H., Shan, S., Nie, Y., & Zhao, Q. (2018). A new boil-off gas re-liquefaction system for LNG carriers based on dual mixed refrigerant cycle. Cryogenics, 92, 84-92. (doi:10.1016/j.cryogenics.2018.04.009)

[83] Ameri, M., & Ahmadi, P. (2007). The study of ambient temperature effects on exergy losses of a heat recovery steam generator. In Challenges of Power Engineering and Environment (pp. 55-60). Springer, Berlin, Heidelberg. (doi:10.1007/978-3-540-76694-0_9)

[84] Szargut, J. (2005). Exergy method: technical and ecological applications (Vol. 18). WIT press.

[85] Mrzljak, V., Poljak, I., & Žarković, B. (2018). Exergy analysis of steam pressure reduction valve in marine propulsion plant on conventional LNG carrier. NAŠE MORE: znanstveno-stručni časopis za more i pomorstvo, 65(1), 24-31. (doi:10.17818/NM/2018/1.4)

[86] Lorencin, I., Anđelić, N., Mrzljak, V., & Car, Z. (2019). Exergy analysis of marine steam turbine labyrinth (gland) seals. Pomorstvo, 33(1), 76-83. (doi:10.31217/p.33.1.8)

[87] Cangioli, F., Chatterton, S., Pennacchi, P., Nettis, L., & Ciuchicchi, L. (2018). Thermo-elasto bulk-flow model for labyrinth seals in steam turbines. Tribology international, 119, 359-371. (doi:10.1016/j.triboint.2017.11.016)

[88] Ameri, M., Mokhtari, H., & Sani, M. M. (2018). 4E analyses and multi-objective optimization of different fuels application for a large combined cycle power plant. Energy, 156, 371-386. (doi:10.1016/j.energy.2018.05.039)

[89] Sengupta, S., Datta, A., & Duttagupta, S. (2007). Exergy analysis of a coal-based 210 MW thermal power plant. International journal of energy research, 31(1), 14-28. (doi:10.1002/er.1224)

[90] Erdem, H. H., Akkaya, A. V., Cetin, B., Dagdas, A., Sevilgen, S. H., Sahin, B., ... & Atas, S. (2009). Comparative energetic and exergetic performance analyses for coal-fired thermal power plants in Turkey. International Journal of Thermal Sciences, 48(11), 2179-2186. (doi:10.1016/j.ijthermalsci.2009.03.007)

[91] Eboh, F. C., Ahlström, P., & Richards, T. (2017). Exergy analysis of solid fuel-fired heat and power plants: a review. Energies, 10(2), 165. (doi:10.3390/en10020165)

[92] Moran, M. J., Shapiro, H. N., Boettner, D. D., & Bailey, M. B. (2010). Fundamentals of engineering thermodynamics. John Wiley & Sons.

[93] Mrzljak, V., Poljak, I., Prpić-Oršić, J., & Jelić, M. (2020). Exergy analysis of marine waste heat recovery $CO_2$ closed-cycle gas turbine system. Pomorstvo, 34(2), 309-322. (doi:10.31217/p.34.2.12)

[94] Wu, J., & Wang, N. (2020). Exploring avoidable carbon emissions by reducing exergy destruction based on advanced exergy analysis: A case study. Energy, 206, 118246. (doi:10.1016/j.energy.2020.118246)

[95] Ibrahim, T. K., Mohammed, M. K., Awad, O. I., Abdalla, A. N., Basrawi, F., Mohammed, M. N., ... & Mamat, R. (2018). A comprehensive review on the exergy analysis of combined cycle power plants. Renewable and Sustainable Energy Reviews, 90, 835-850. (doi:10.1016/j.rser.2018.03.072)

[96] Ali, M. S., Shafique, Q. N., Kumar, D., Kumar, S., & Kumar, S. (2020). Energy and exergy analysis of a 747-MW combined cycle power plant Guddu. International Journal of Ambient Energy, 41(13), 1495-1504. (doi:10.1080/01430750.2018.1517680)

[97] Wang, C., Yan, C., Wang, J., Tian, C., & Yu, S. (2017). Parametric optimization of steam cycle in PWR nuclear power plant using improved genetic-simplex algorithm. Applied Thermal Engineering, 125, 830-845. (doi:10.1016/j.applthermaleng.2017.07.045)

[98] Mrzljak, V., Lorencin, I., Anđelić, N., & Car, Z. (2021). Thermodynamic Analysis of a Condensate Heating System from a Marine Steam Propulsion Plant with Steam Reheating. Journal of Marine Science and Application, 1-11. (doi:10.1007/s11804-021-00191-5)

[99] Lemmon, E. W., Huber, M. L., & McLinden, M. O. (2010). NIST Standard Reference Database 23, Reference Fluid Thermodynamic and Transport Properties (REFPROP), version 9.0, National Institute of Standards and Technology. R1234yf. fld file dated December, 22, 2010.

[100] Ibrahim, T. K., Basrawi, F., Awad, O. I., Abdullah, A. N., Najafi, G., Mamat, R., & Hagos, F. Y. (2017). Thermal performance of gas turbine power plant based on exergy analysis. Applied thermal engineering, 115, 977-985. (doi:10.1016/j.applthermaleng.2017.01.032)

[101] Ehsana, A., & Yilmazoglu, Z. (2011). Design and exergy analysis of a thermal power plant using different types of Turkish lignite. International Journal of Thermodynamics, 14(3), 125-133. (doi:10.5541/ijot.288)

[102] Tontu, M., Sahin, B., & Bilgili, M. (2019). Using energy and exergy analysis to compare different coal-fired power plants. Energy Sources, Part A: Recovery, Utilization, and Environmental Effects, 1-16. (doi:10.1080/15567036.2019.1696429)

[103] Sayyaadi, H., & Sabzaligol, T. (2009). Various approaches in optimization of a typical pressurized water reactor power plant. Applied Energy, 86(7-8), 1301-1310. (doi:10.1016/j.apenergy.2008.10.011)

[104] Mitrović, D., Zivkovic, D., & Laković, M. S. (2010). Energy and exergy analysis of a 348.5 MW steam power plant. Energy Sources, Part A: Recovery, Utilization, and Environmental Effects, 32(11), 1016-1027. (doi:10.1080/15567030903097012)

[105] Poljak, I., Bielić, T., Mrzljak, V., & Orović, J. (2020). Analysis and Optimization of Atmospheric Drain Tank of Lng Carrier Steam Power Plant. Journal of Marine Science and Engineering, 8(8), 568. (doi:10.3390/jmse8080568)

[106] Anđelić, N., Baressi Šegota, S., Lorencin, I., & Car, Z. (2020). Estimation of gas turbine shaft torque and fuel flow of a CODLAG propulsion system using genetic programming algorithm. Pomorstvo, 34(2), 323-337. (doi:10.31217/p.34.2.13)

[107] Lorencin, I., Anđelić, N., Mrzljak, V., & Car, Z. (2019). Multilayer Perceptron approach to Condition-Based Maintenance of Marine CODLAG Propulsion System Components. Pomorstvo, 33(2), 181-190. (doi:10.31217/p.33.2.8)

[108] Baressi Šegota, S., Lorencin, I., Musulin, J., Štifanić, D., & Car, Z. (2020). Frigate Speed Estimation Using CODLAG Propulsion System Parameters and Multilayer Perceptron. NAŠE MORE: znanstveni časopis za more i pomorstvo, 67(2), 117-125. (doi:10.17818/NM/2020/2.4)

[109] Anđelić, N., Baressi Šegota, S., Lorencin, I., Mrzljak, V., & Car, Z. (2021). Estimation of COVID-19 epidemic curves using genetic programming algorithm. Health Informatics Journal, 27(1), 1460458220976728. (doi:10.1177/1460458220976728)